 \documentclass[12pt,preprint]{aastex}

\shorttitle{The shortest modulation period Blazhko star: SS Cnc}
\shortauthors{Jurcsik et al.}

\begin{document}

\title{The shortest modulation period Blazhko RR Lyrae star: SS Cnc}

\author{J. Jurcsik, B. Szeidl, and \'A. S\'odor,}
\affil{Konkoly Observatory, P.O.~Box~67,
H-1525 Budapest, Hungary, \email{jurcsik@konkoly.hu}}

\author{I. D\'ek\'any, Zs. Hurta, K. Posztob\'anyi, and K. Vida,}
\affil{E\"otv\"os Lor\'and University, Dept. of Astronomy, P.O.~Box
~32, H$-$1518
Budapest, Hungary}

\author{M. V\'aradi}

\affil{Konkoly Observatory,  P.O.~Box~67,
H$-$1525 Budapest, Hungary}

\author{A. Szing}

\affil{Dept. of Experimental Physics and Astronomical Observatory, 
University of Szeged, Hungary}

\begin{abstract}
Extended $BV(RI)_C$ CCD observations of SS Cnc, a short period RRab star 
are presented. Nearly 1400 data points in each band have been obtained 
spanning over 79 days during the spring of 2005.
The star exhibits light curve modulation, the so called Blazhko effect
with small amplitude ($B$ maximum brightness varies 0.1 mag) and
with the shortest modulation period (5.309~d) ever observed. 
In the Fourier spectrum of the $V$ light curve the pulsation frequency 
components are detected up to the 24th harmonic order, and modulation 
side lobe frequencies with significantly asymmetric amplitudes are seen 
up to the 15th and 9th orders for the lower and higher frequency
components, respectively. Detailed comparison of the modulation behavior of  
SS Cnc and RR Gem, the two recently discovered small amplitude, short 
modulation period Blazhko stars is presented. The modulation frequency 
($f_m$) appears in the Fourier spectrum of both stars with similar amplitude.  
We also demonstrate that the modulation frequencies have basically different
properties as the pulsation and modulation side lobe frequencies have, 
indicating that the physics behind these frequency components are not the same.
The discovery of small amplitude modulations of RRab stars cautions that 
the large photometric surveys (MACHO, OGLE) may  seriously underestimate 
the number of modulated RR Lyrae stars.

\end{abstract}


\section{Introduction}

RR Lyrae stars are of great astrophysical importance in many respects.
They are standard candles of distance estimations, test objects of stellar
evolution and pulsation model calculations. The physics governing their
pulsation behavior  is thought to be well understood, but there are still 
some puzzling properties of these objects as well. The most troublesome is 
the so called Blazhko effect, the light curve modulation of some fraction 
of RR Lyrae stars. 
\citet{smith} gives an overall description of the phenomenon, and for a 
recent summary of the problems of theoretical explanations see \citet{dm}.
First connections between some properties of the pulsation and the modulation 
have been established in \citet{acta} and \citet{ibvs}.

The Blazhko phenomenon has been known for about a century. Most of the
known galactic field RRab stars showing light curve modulation were discovered 
from photographic observations during the first part of the last century.
Though the advent of the photoelectric, and in the last decades of the 20.
century, the CCD era has led to spectacular increase of the photometric precision,
there is still a lack of extended high accuracy survey of the modulation 
properties of the galactic field RR Lyrae stars in the Solar neighborhood.

The statistics of the occurrence frequency of the phenomenon are contradictory, 
e.g. \citet{mp} found 35\% and 5\% incidence rate for fundamental and first overtone
variables in the Galactic Bulge, while \citet{machoab} and \citet{macho1} detected 
11.9\% and  7.5\% of the RR Lyrae stars to show modulation properties in the LMC, 
respectively. Because of the differences in the accuracy of the surveys, the 
photometric wavebands used, the extension of the datasets, and the analyzing  
methods applied, exact comparison of the different statistics is, however, 
questionable. 

The recent discovery of the very small amplitude modulation of RR Geminorum 
\citep{rrg} called our attention to that small amplitude modulations which have 
escaped notice in the previous photometric studies might be quite frequent. 
This possibility may override any previous statistics of the incidence rate of 
the modulation. In order to get a clear picture of the frequency of the modulation 
of galactic field RR Lyrae stars, we started a systematic photometric survey 
at Konkoly Observatory to measure short period ($P<0.48$ d) fundamental mode 
RR Lyrae stars.

In this note we report the discovery of another RR Lyrae star, SS Cnc 
($\alpha=08^h 06^m 25\fs56, \delta=+23\degr 15\arcmin 05\farcs8$, J2000), 
exhibiting small amplitude modulation. Previous photometric observations of 
SS Cnc are very sparse, altogether 24 accurate photometric data points have 
been published by different authors \citep{fitch,sturch,epstein}.
It is a metal rich RR Lyrae star, its [Fe/H]$=-0.03$
\citep[see the metallicity compilation in][]{jk96}.

\section{Observations and data reduction}

The observations were obtained with the automated 60cm telescope of Konkoly
Observatory, Sv\'abhegy, Budapest equipped with a Wright 750x1100 CCD 
using $BV(RI)_C$ filters. The field of view was 17'x24'. Data reduction was 
performed using standard {\sc 
IRAF}\footnotemark[1]\footnotetext[1]{{\sc IRAF} 
is distributed by the National Optical Astronomy Observatories,
    which are operated by the Association of Universities for Research
    in Astronomy, Inc., under cooperative agreement with the National
    Science Foundation.} packages.

Nearly 1400 frames were obtained in each passband on 35 nights
during the spring of 2005 (between JD~2\,453\,388 and JD~2\,453\,467).
The data were transformed to standard $B, V, R_C, I_C$ magnitudes 
by using observations of standard stars in M67 \citep{m67}. 

As no single comparison star with ideal properties (brightness, color, vicinity)
has been found in the field we have measured relative magnitudes of SS Cnc 
to the mean magnitudes of three stars: GSC 01927-00504, GSC 01927-00744, 
and GSC 01927-00835. 

The $r.m.s.$ scatter of the averaged magnitude differences between these 
three stars is 0.008, 0.009, 0.009, and 0.009 mag in the $B, V, R_C, I_C$ 
colors, respectively.

Relative magnitudes of SS Cnc are available in machine readable format
at the http:/www.konkoly.hu/24/publications/ web address.  

\section{The shortest modulation period, 5.309 day, ever observed}

\begin{figure}
\epsscale{.80}
\plotone{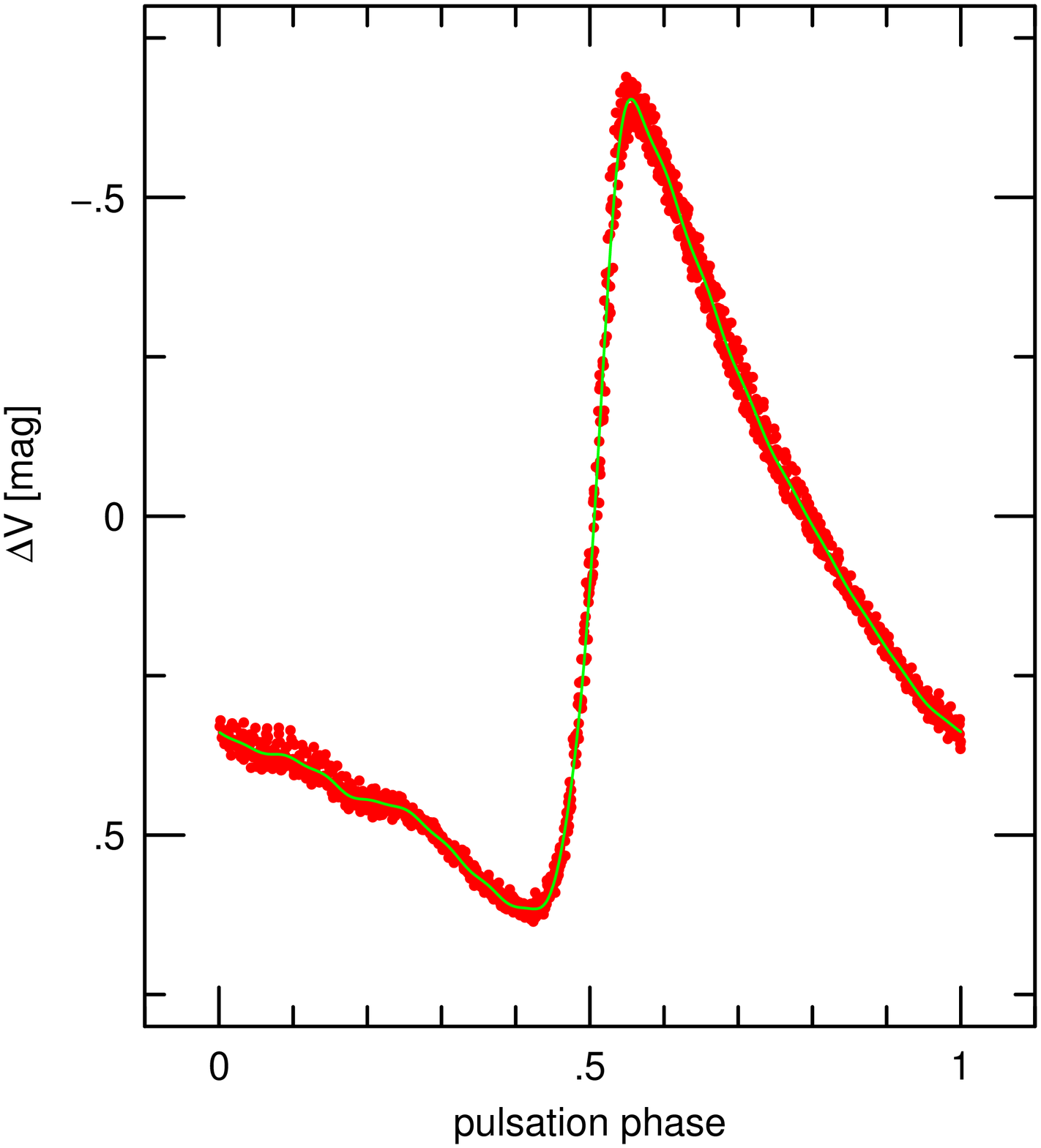}
\caption{$\Delta V$ light curve of SS Cnc folded with the 0.367337 d pulsation period. 
The line corresponds to a fit with the pulsation frequency and its 24 harmonic 
components. The 0.05--0.10 mag width of the light curve is due to modulation 
with an extremely short, 5.309 d period, and of small amplitude. 
Phase 0.5 corresponds to the mid of the rising branch, where
the measured magnitude equals to the intensity mean magnitude. 
\label{figlc}}
\end{figure}

The observed pulsation light curve of SS Cnc (see Fig.~\ref{figlc}) 
shows larger scatter than observational inaccuracies could explain.
Fitting the light curve only with the pulsation frequency and its harmonics 
the $r.m.s.$ scatter of the residual data is 2-3 times larger than expected, 
0.028, 0.022, 0.017 and 0.014 mag in the $BVR_CI_C$ colors, respectively. 

\begin{figure}
\epsscale{1.10}
\plottwo{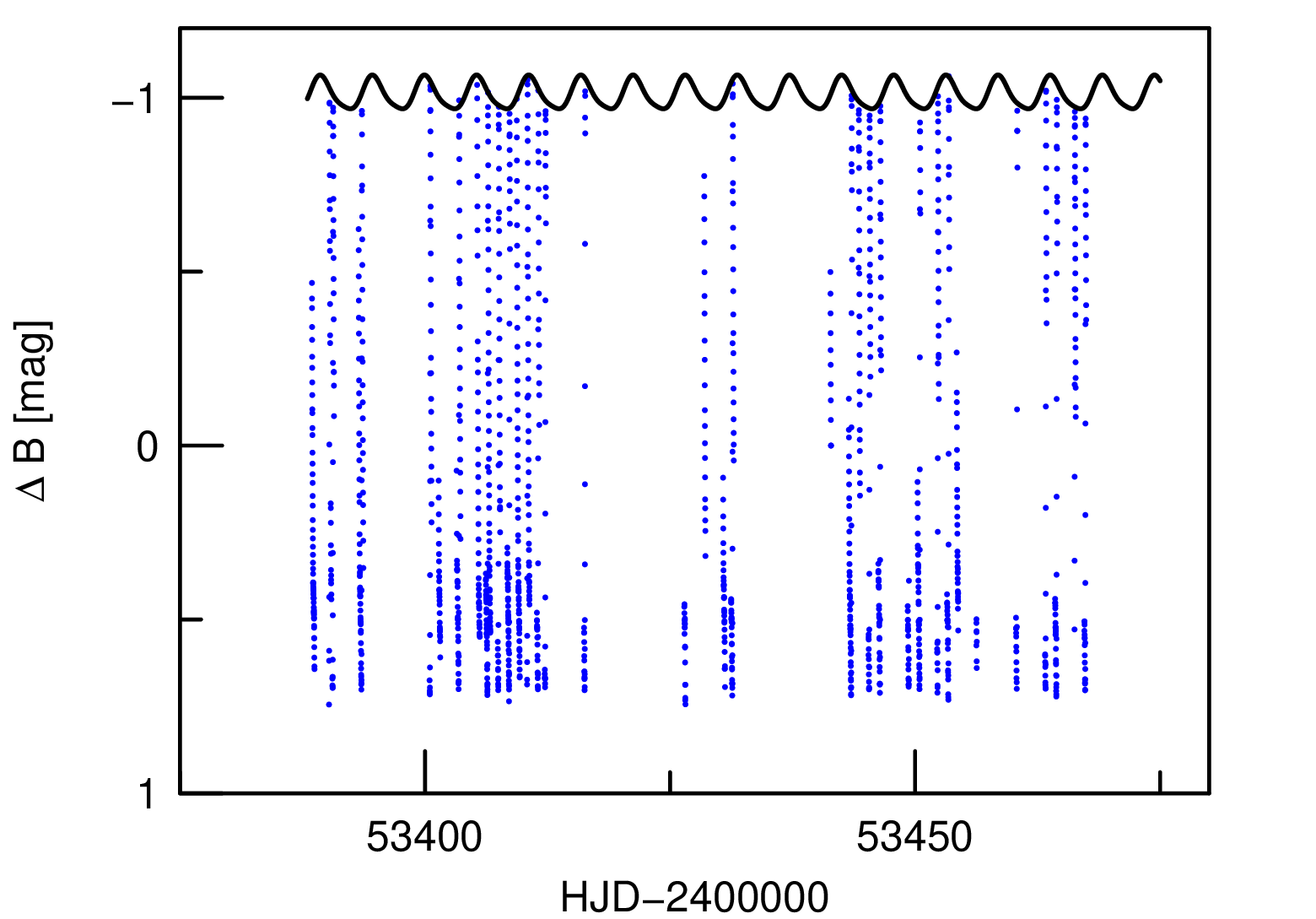}{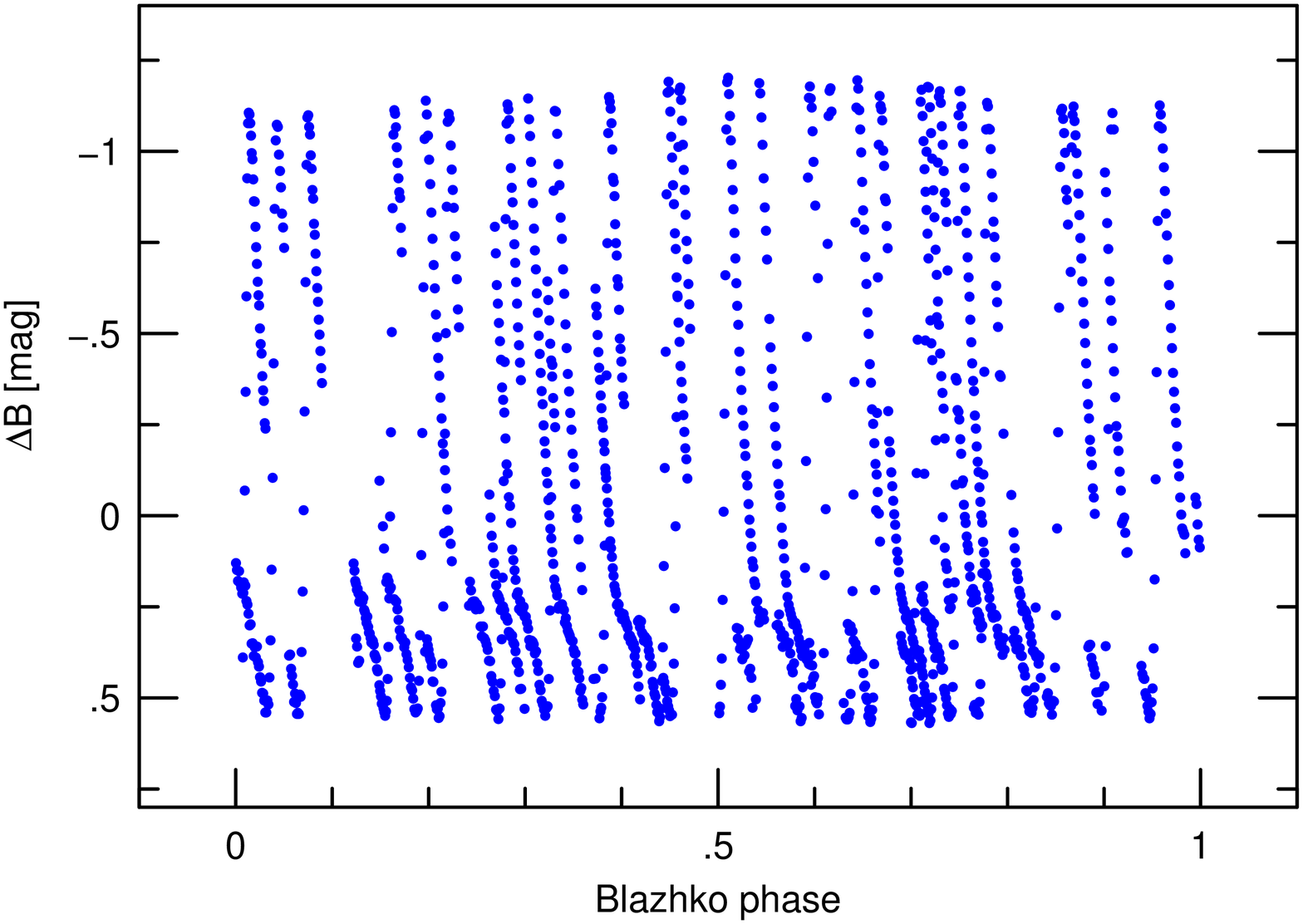}
\caption{$\Delta B$ observations of SS Cnc (left panel) and data folded according 
to the phase of the detected 5.309 d modulation period (right panel).
Due to the shortness of the pulsation period data cover maximum light
in most of the nights of the observations. Therefore, the upper envelope of all 
the measurements shown in the left panel indicates clearly modulation with
5.309 d and total amplitude of 0.10 mag. The asymmetric shape of the envelope 
is a sign that beside amplitude modulation phase modulation is also present.
\label{figmind}}
\end{figure}

The brightness of maximum light varied from night to night, indicating
Blazhko modulation of the light curve. Detailed Fourier analysis of the data
has revealed that this is indeed the case, but the modulation period of SS Cnc 
is unexpectedly short, only 5.309 d. In Fig.~\ref{figmind} $\Delta B$ 
observations are plotted against HJD and  the phase of the detected short 
period modulation. 

\begin{figure}
\epsscale{.70}
\plotone{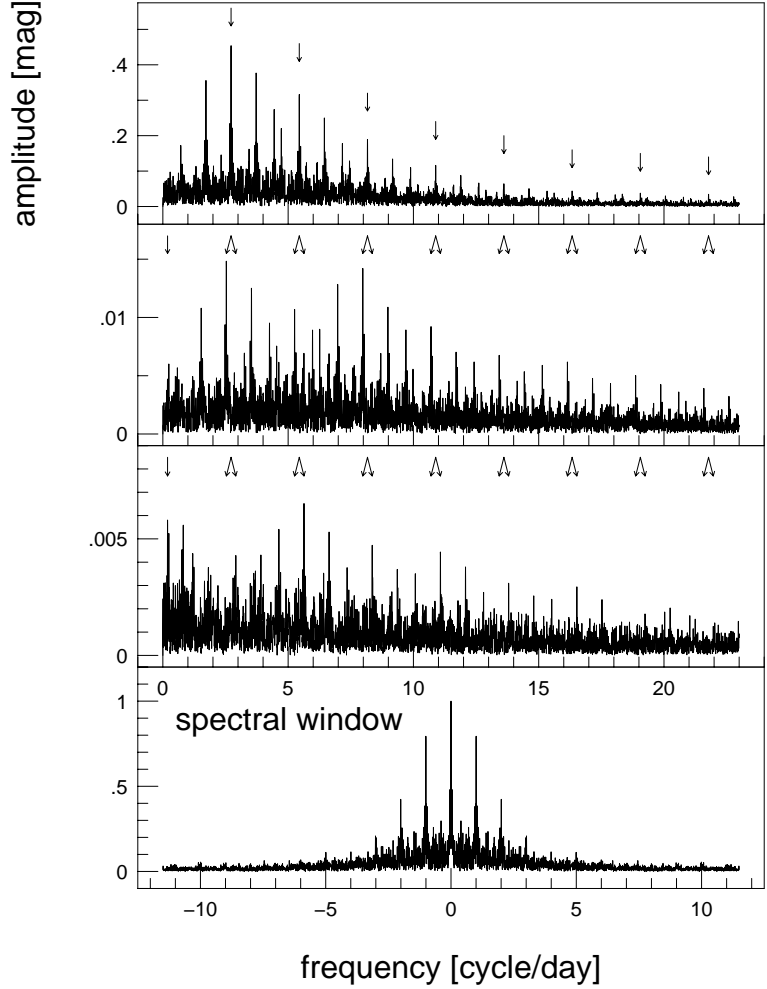}
\caption{Fourier spectrum of the $V$ light curve of SS Cnc. 
In the top panel arrows point to the pulsation frequency and its harmonics ($kf_0$). 
The other large amplitude peaks are alias frequencies of these components.
The next panel shows the prewhitened spectrum after the removal of the signal 
of the radial mode pulsation. In this spectrum modulation frequency components 
appear at 0.18835 c/d shorter positions than the pulsation frequencies ($kf_0-f_m$).
Arrows show modulation frequency positions at the sides of equidistant triplets 
($kf_0 \pm f_m$) appearing typically in the spectra of Blazhko stars. 
The larger frequency side lobe modulation components ($kf_0+f_m$) can be detected
only after the $kf_0 - f_m$ components have been removed as shown in the next panel.
It is apparent that the amplitudes of the modulation side lobe frequencies  are 
strongly asymmetric, among the longer frequency components only $2f_0+f_m$ has 
an amplitude larger than 0.005 mag. The separations of the modulation frequency components 
correspond to 5.309 d period of the modulation. Bottom panel shows the spectral window.
\label{figsp}}
\end{figure}

The pulsation and modulation periods (mean values obtained for the 
$BVR_CI_C$ data) have been determined from Fourier analysis, using the 
facilities of the program package MUFRAN \citep{mufran}. We have detected 
24 harmonics of the pulsation frequency ($f_0=2.722296$ c/d) in the Fourier 
spectrum of the $V$ light curve, and modulation peaks up to the 15th and 9th 
orders at smaller and larger frequency values than the pulsation components, 
respectively (see Fig.~\ref{figsp}). The positions of the side lobe modulation 
frequencies ($kf_0 \pm f_m$) correspond to $f_m=0.18835$ c/d, i.e., 5.309 d 
period of the modulation. The following ephemerides give maximum pulsation 
light and maximum amplitude timings:

HJD $T_{max} = 2\,453\,464.459 + 0.367337 E_{pulsation}$,

HJD $T_{Amax}= 2\,453\,410.6 + 5.309 E_{Blazhko}$.

The amplitudes ($A_V$) and phases ($\phi_V$) of the detected frequencies of the 
$V$ light curve and $A_B/A_V, A_V/A_R$ and $A_V/A_I$ amplitude ratios and 
$\phi_B-\phi_V, \phi_V-\phi_R$ and $\phi_V-\phi_I$ phase differences are listed 
in Table~\ref{tablefour}. The phases correspond to sine term solutions with initial
epoch value: $T_0=2\,453\,388.000$.

The amplitudes of the modulation side lobe frequencies ($kf_0 \pm f_m$) are strongly
asymmetric, the smaller frequency components ($kf_0-f_m$) have significantly
larger amplitudes than the longer frequency ones ($kf_0+f_m$).
With the exception of the second order, the amplitudes of the $kf_0+f_m$
components do not exceed 0.005 mag in $V$ band. This order of magnitude signal 
definitely escapes detection in the case of less extended datasets and/or 
worse S/N ratio. 

This result warns that the classification scheme of the modulation which is 
based on that the modulation side lobe frequencies are present only at one side 
of the pulsation components ($\nu1$ variables) or at both sides (Blazhko variables)  
might be seriously affected by data incompleteness and S/N properties.
A similar conclusion has already been drawn in \citet{rrg} based on 
that the Fourier spectra of the two halves of the extended dataset of RR Gem
show substantially different character of the modulation.
According to the data sampling and accuracy of the existing photometric 
surveys, SS Cnc would have been classified as a $\nu1$  variable
if its modulation had been detected at all.

\section{Comparison of the modulation properties of the two small modulation  
amplitude Blazhko stars, SS Cnc and RR Gem}

Our multicolor photometric observations of  SS Cnc and RR Gem \citep{rrg}
are extended enough to study and to compare the modulation characteristics of 
these stars in detail. Though both stars exhibit modulation with extremely 
short period and small amplitude their modulation properties are different in 
many respects.

\begin{figure}
\epsscale{1.1}
\plottwo{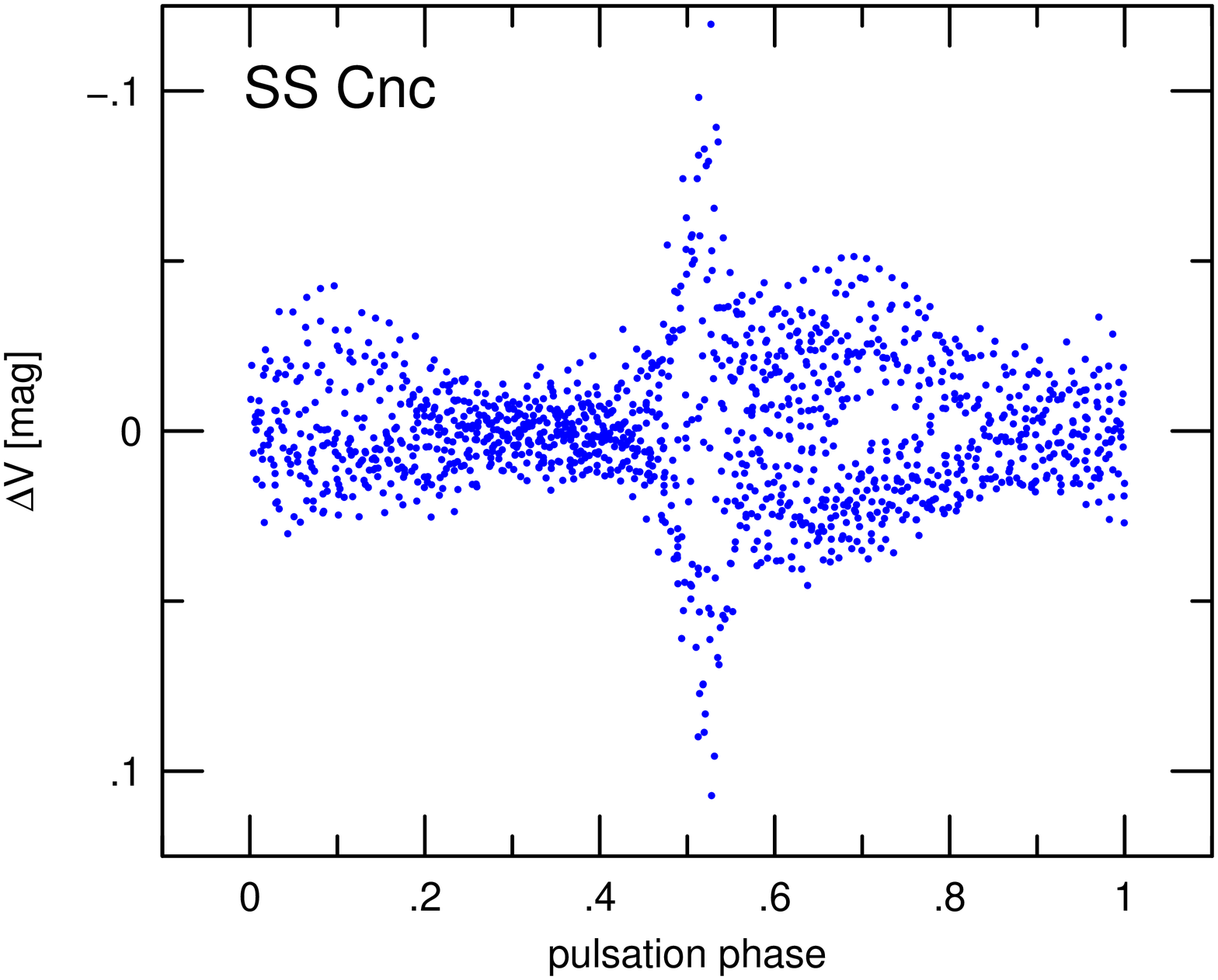}{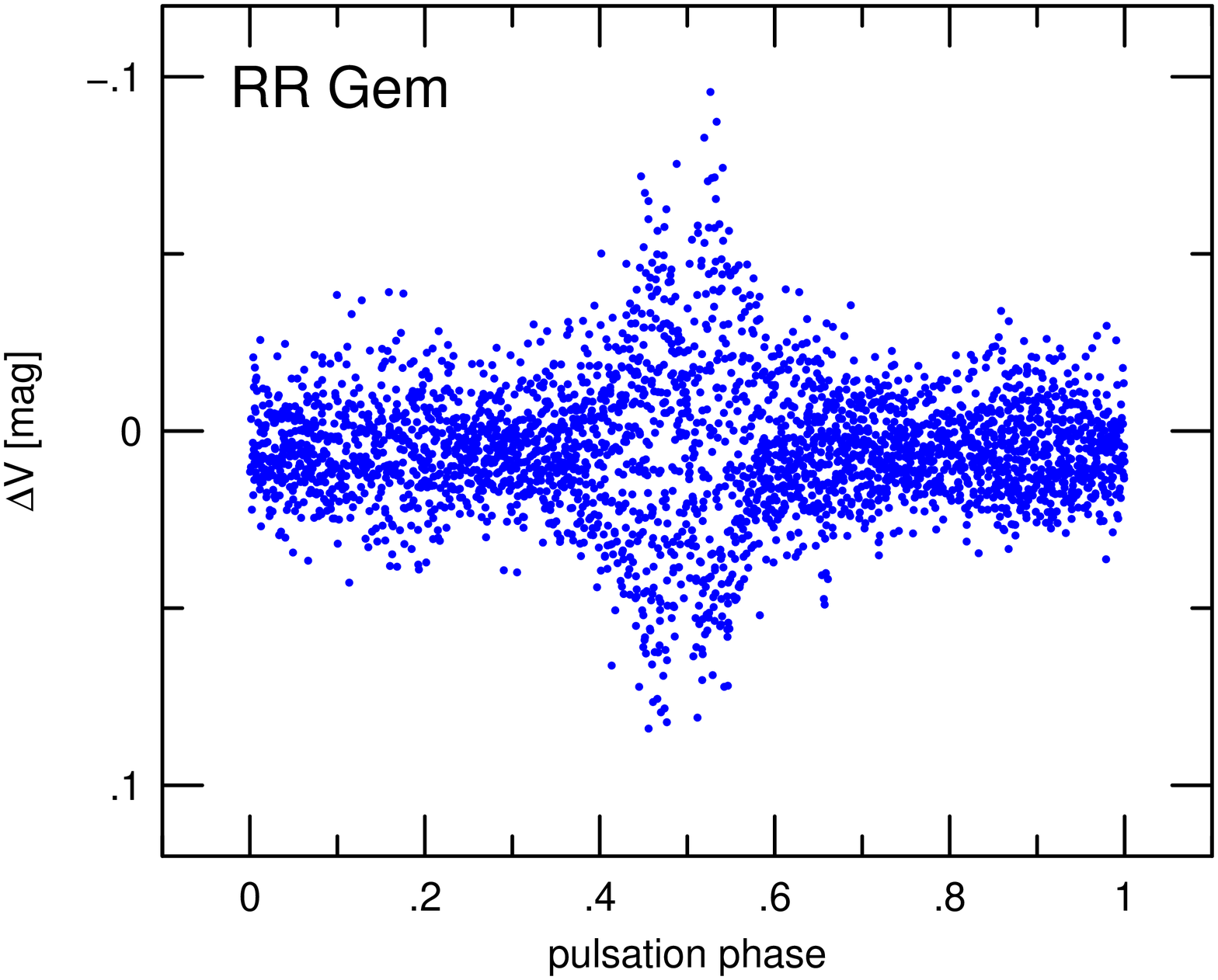}
\caption{Residual light curves of SS Cnc (left) and RR Gem (right) after the
removal of the mean light variation according to a Fourier fit to the data  
with the pulsation frequency and its harmonics. In both plots phase 0.5 corresponds
to the mid of the rising branch. In SS Cnc some hundredths of magnitude modulation 
can be seen in most phases of the pulsation, except between the phases 0.25--0.45 
which falls into the 0.2 phase interval preceding minimum light (see Fig.~\ref{figlc}). 
The amplitude of the modulation is the largest on the rising branch, in the phase 
interval 0.47--0.55. The residual light curve of RR Gem is different. Its symmetry 
to phase 0.5 is evident, enhanced modulation occurs in two lobes in 0.1 phase 
intervals preceding and following phase 0.5. In SS Cnc the left side lobe of the 
residual curve seems to be missing.
\label{figres}}
\end{figure}

Removing the mean pulsation light curves from the $V$ data the residual light 
curves against the pulsation phases are shown in Fig.~\ref{figres} for both stars. 
These plots indicate only slight similarity, while the residual plot of RR Gem is
symmetrical to phase 0.5 (which corresponds to the mid of the rising branch in both
plots), there is no symmetry in the residual data of SS Cnc. The only similarity of 
the two plots is that large modulation amplitude is restricted to a narrow phase 
interval in the vicinity of phase 0.5. Similar residual phase plot of DR And, 
a large modulation amplitude Blazhko star was shown in \citet{drand}. The modulation 
is the largest on the ascending branch of DR And, too, but increased modulation
extends to a wider, 0.4 pulsation phase interval.  

The residual light curves of the Blazhko variables give some impression about 
the character of the modulation. In case of pure amplitude modulation the residual 
is the largest at around minimum and maximum light and it is the smallest at around
mean brightness values. On the contrary, pure phase modulation results in large 
residual amplitude at around mid brightnesses and small residual amplitudes around 
the extrema. Based on these properties Fig.~\ref{figres} indicates that in SS Cnc 
phase modulation does also occur, while in RR Gem the amplitude modulation is dominant.

The data coverage enabled us to analyze the displacement of the extrema and the shape 
of the light curves in the different phases of the Blazhko cycle separately. 
In order to do this, we have divided the $V$ dataset of the two stars 
into ten-ten subsets, each containing data of a full pulsation light curve 
but in different phase intervals of the Blazhko cycle. (In SS Cnc there is a 
0.2 pulsation phase gap in one of these light curves which was filled by linear
interpolation of the data  subsets of the preceding and following Blazhko phases.)

\begin{figure}
\epsscale{1.}
\plotone{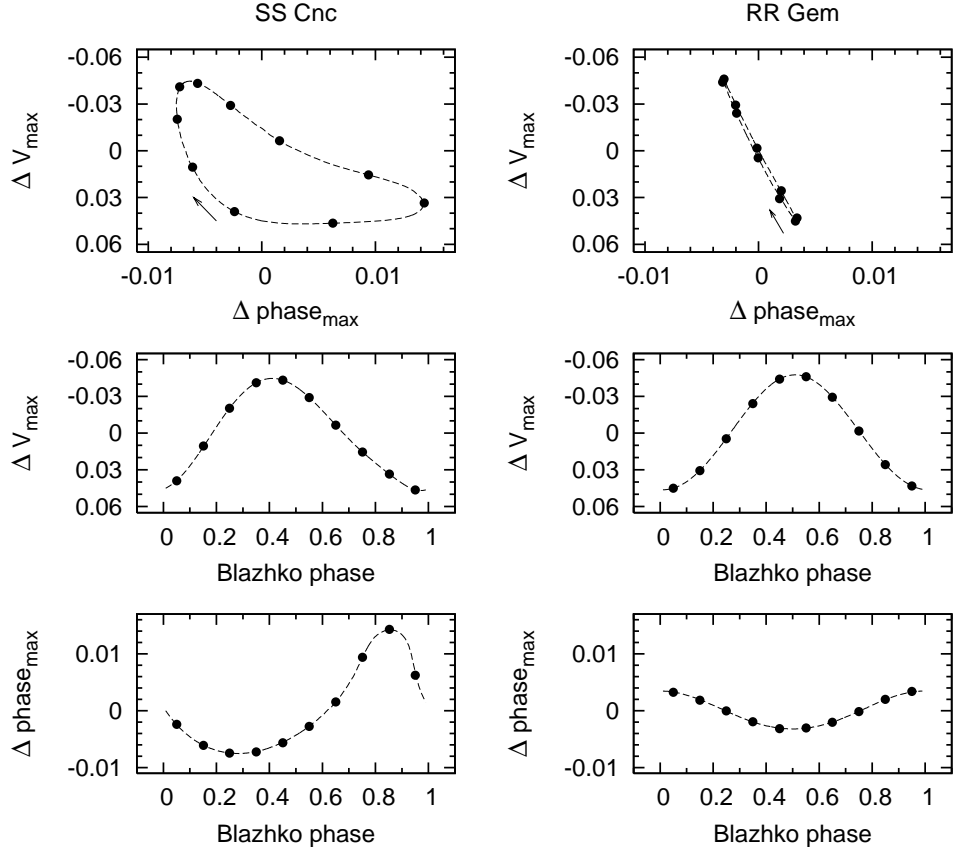}
\caption{
Variations in the heights and phases of maxima of SS Cnc and RR Gem during the 
Blazhko cycle are plotted. $\Delta V_{max}$ and $\Delta phase_{max}$ denote $V_{max} -
\langle V_{max} \rangle$ [mag] and $phase_{max}-\langle phase_{max} \rangle$ [d],
respectively. Both stars exhibit amplitude and phase modulations, but in RR Gem  
the amplitude modulation is by far dominant. In the top figures arrows show the 
amplitude and phase variations between the first two Blazhko phases, the progression 
during the Blazhko cycle is clockwise in both stars.
\label{amp-phase}}
\end{figure}

The variations in the heights and phases of the maxima in 10 different Blazhko phases 
are shown in Fig.~\ref{amp-phase} for the two stars. From these plots it is evident 
that in both stars both amplitude and phase modulation occur, but, indeed,
in RR Gem amplitude modulation is dominant. Maxima timings precede the expected value   
the most at around the largest amplitude phase of the Blazhko cycle both in
SS Cnc and in RR Gem.

\begin{figure}
\epsscale{0.8}
\plotone{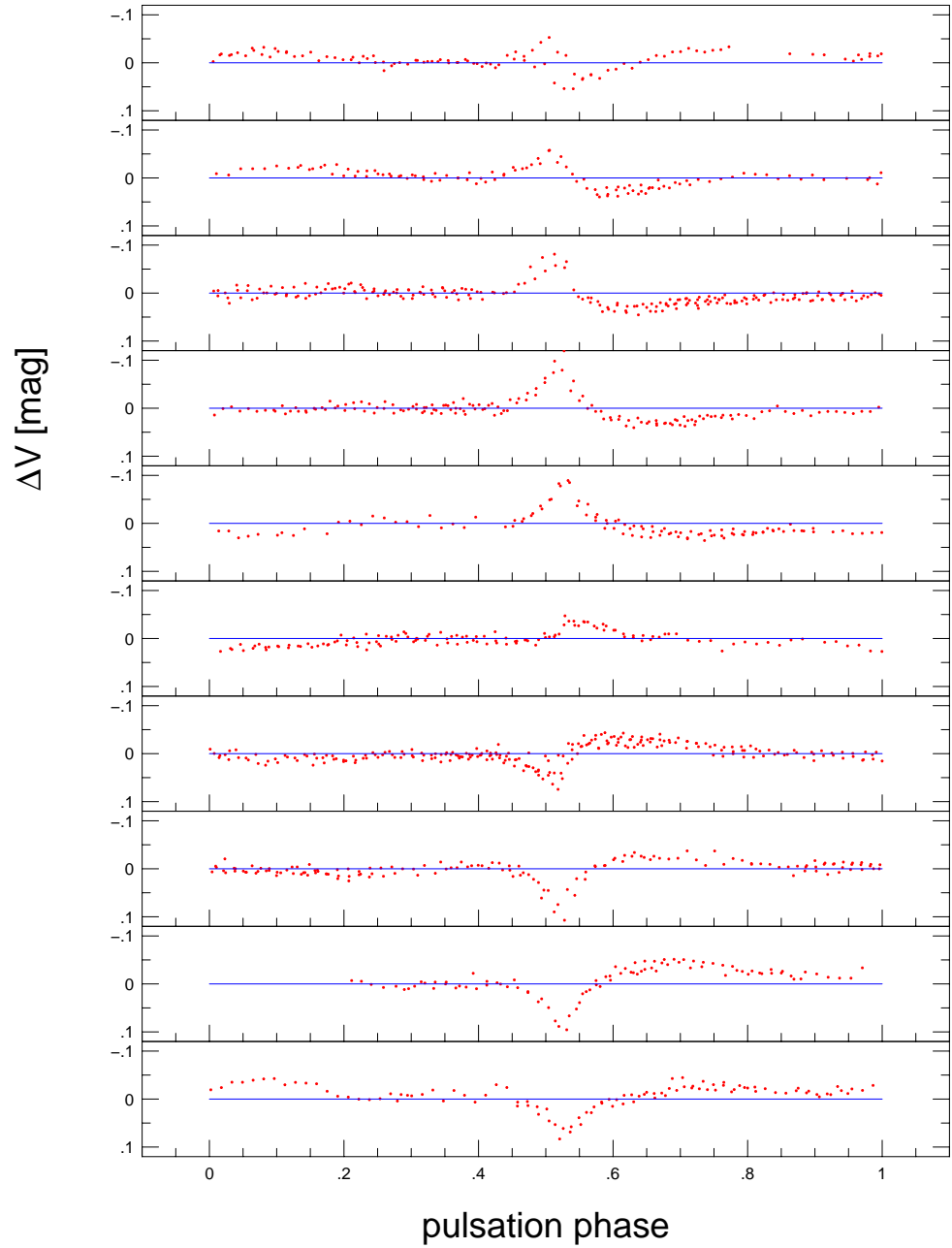}
\caption{ Residual light curves of SS Cnc belonging to 0.1 Blazhko phase 
intervals are plotted separately. 
\label{figres10}}
\end{figure}

\begin{figure}
\epsscale{1.0}
\plottwo{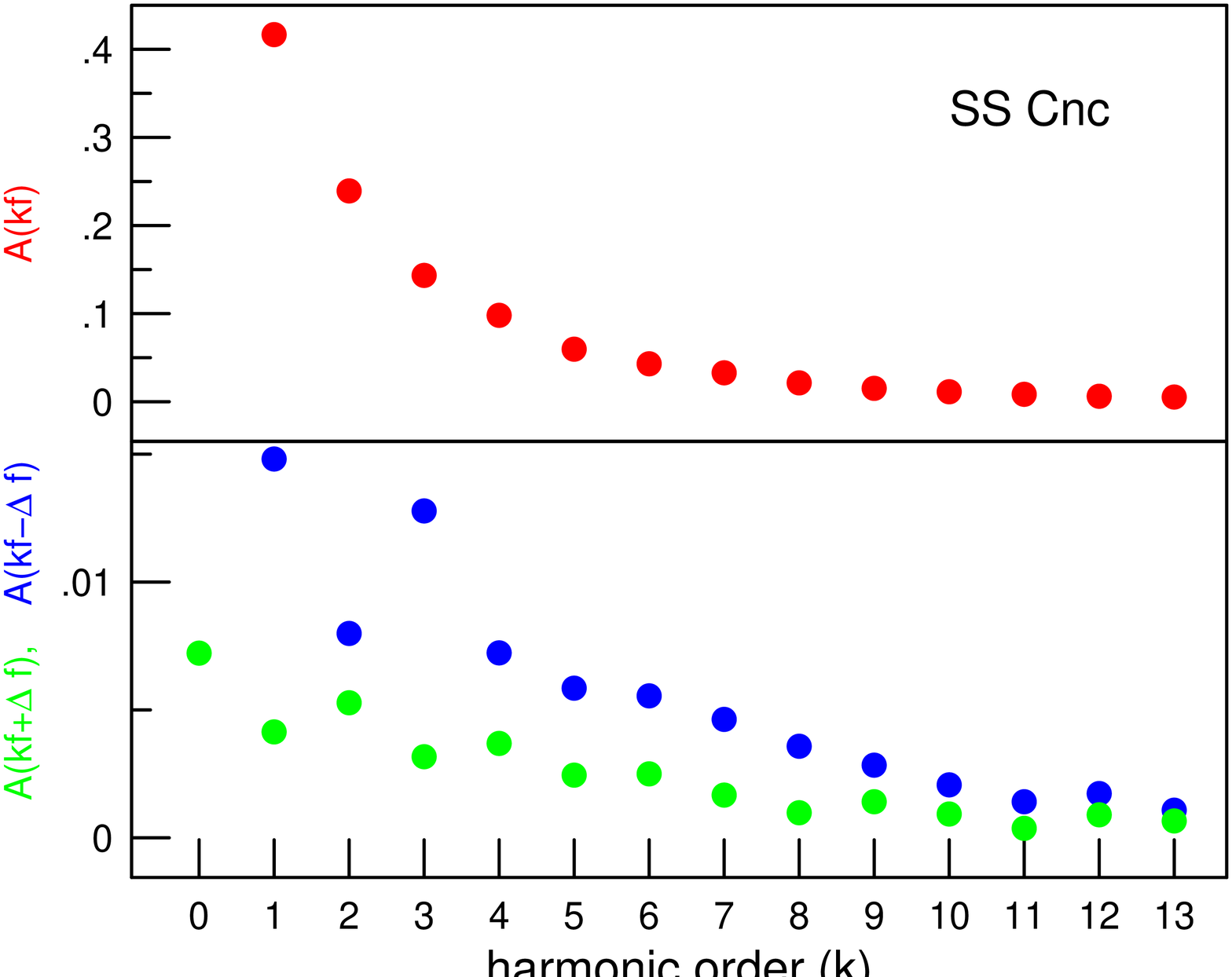}{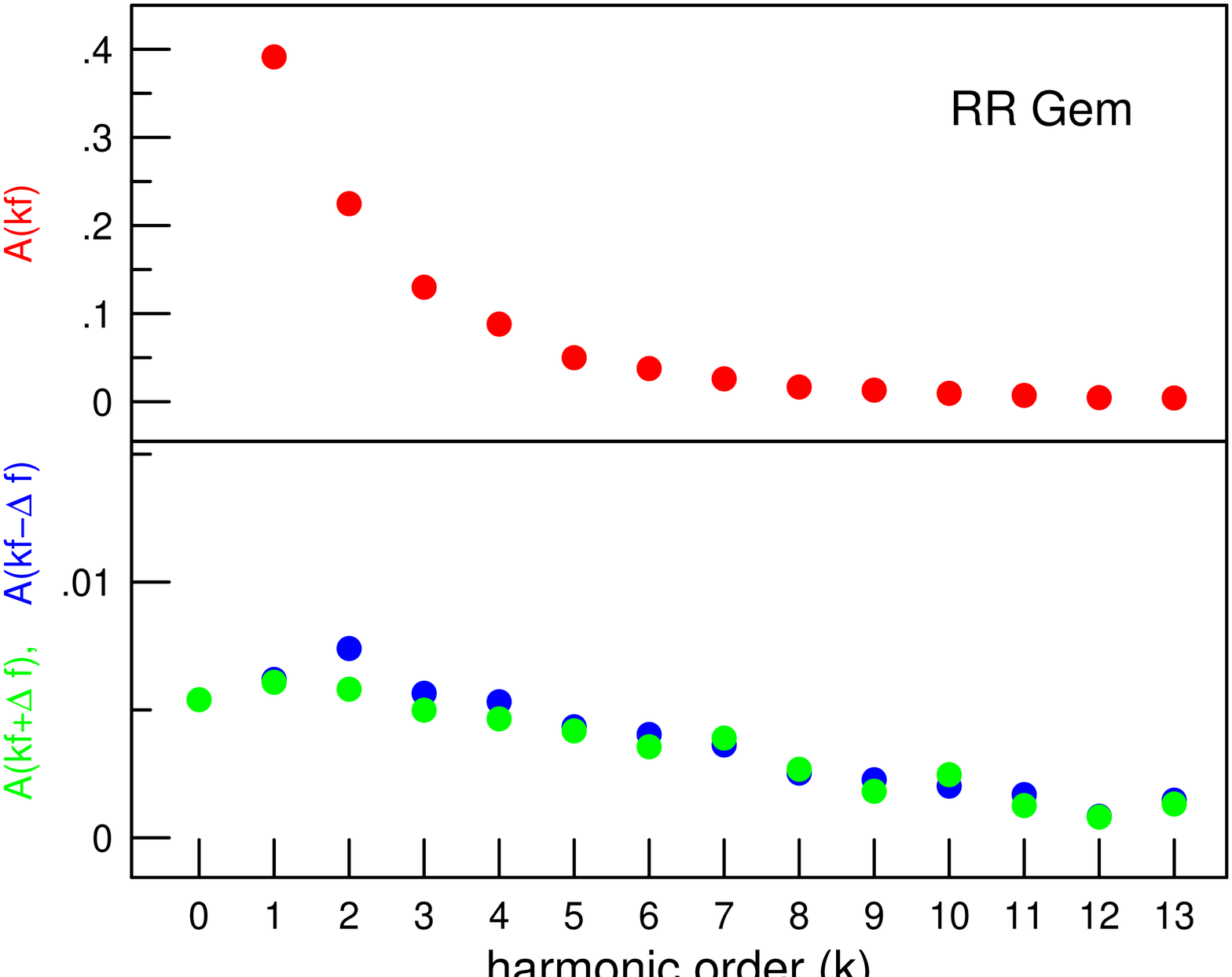}
\caption{Fourier  amplitudes  of the pulsation and modulation frequencies 
for SS Cnc and RR Gem in $V$ band. In both cases the amplitudes of the pulsation 
frequencies decrease exponentially towards higher orders, while the amplitudes 
of the modulation frequencies decrease roughly linearly with some irregularity.
The $kf_0-f_m$ and $kf_0+f_m$ modulation components of SS Cnc have significantly 
different, asymmetric amplitudes while for RR Gem they have similar amplitudes. 
It is an interesting feature, that in both figures the amplitudes of the second
order modulation components show different character as the first and third
order ones, it is more symmetric in SS Cnc and more asymmetric in RR Gem than 
the amplitudes of the neighboring order modulation frequencies.
\label{figamp}}
\end{figure}

In Fig.~\ref{figres10} the residual light curves of SS Cnc in the 10 different 
phases of the Blazhko cycle are separately plotted. Again, it is evident that 
large amplitude modulation occurs in a restricted phase interval around phase 0.5.
This concentration of the modulation to a narrow phase interval of the pulsation 
results in enhanced amplitudes of the higher order modulation components.
Fig.~\ref{figamp} compares the amplitude decrease of the pulsation and modulation
frequency components with increasing harmonic orders of the two stars. In contrast 
to the exponential amplitude decrease of the pulsation components, the amplitudes 
of the modulation frequencies decrease less steeply. Another interesting feature 
is that, though the $kf+f_m$ and  $kf-f_m$ components have the same amplitudes 
in RR Gem and are significantly different in SS Cnc, the mean values of the 
amplitudes at a given order are very similar in the two stars. It is also important 
to note that $f_m$ appears with commensurable amplitude like the first modulation
frequencies in both cases.

\begin{figure}
\epsscale{1.1}
\plottwo{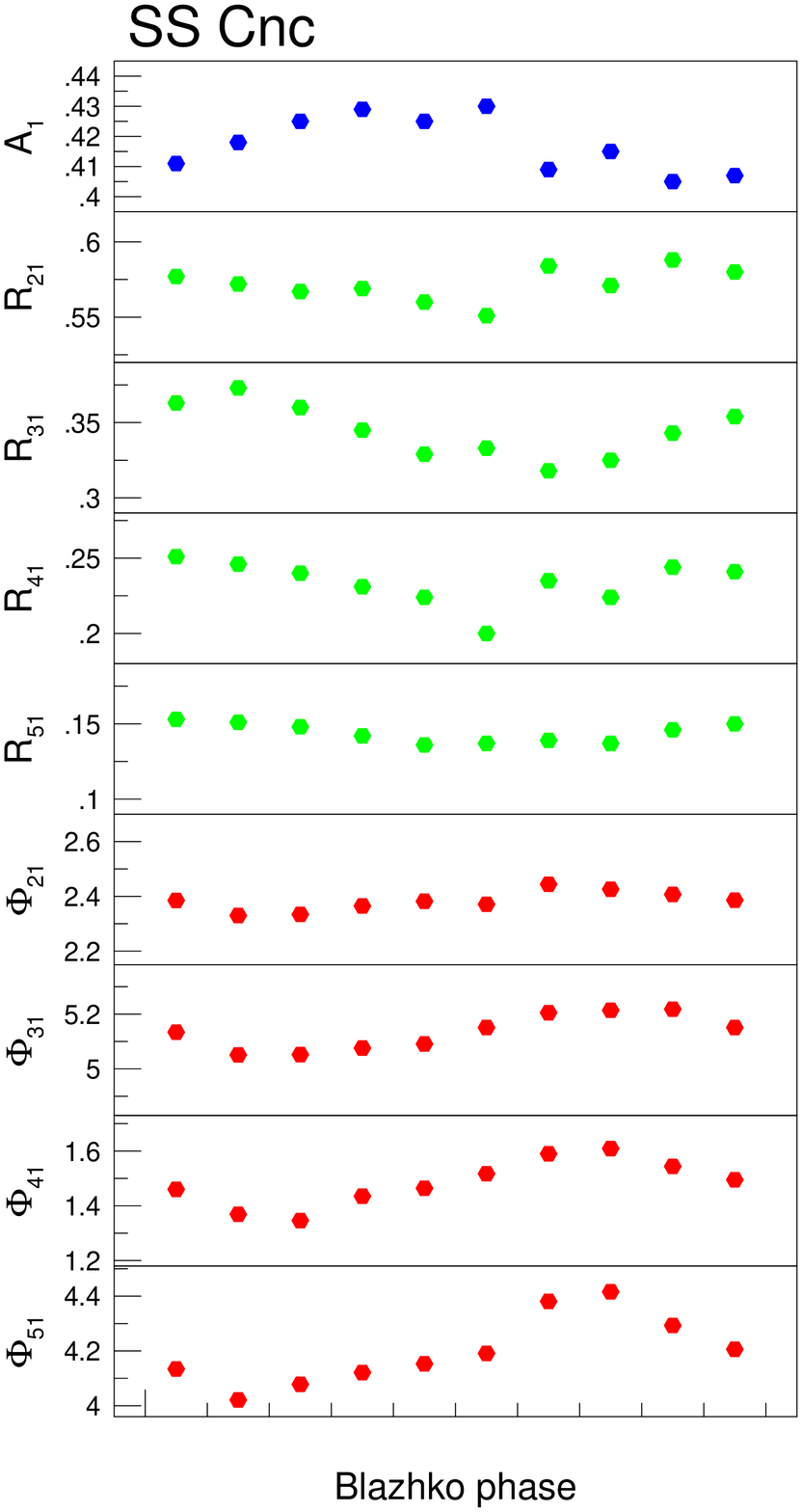}{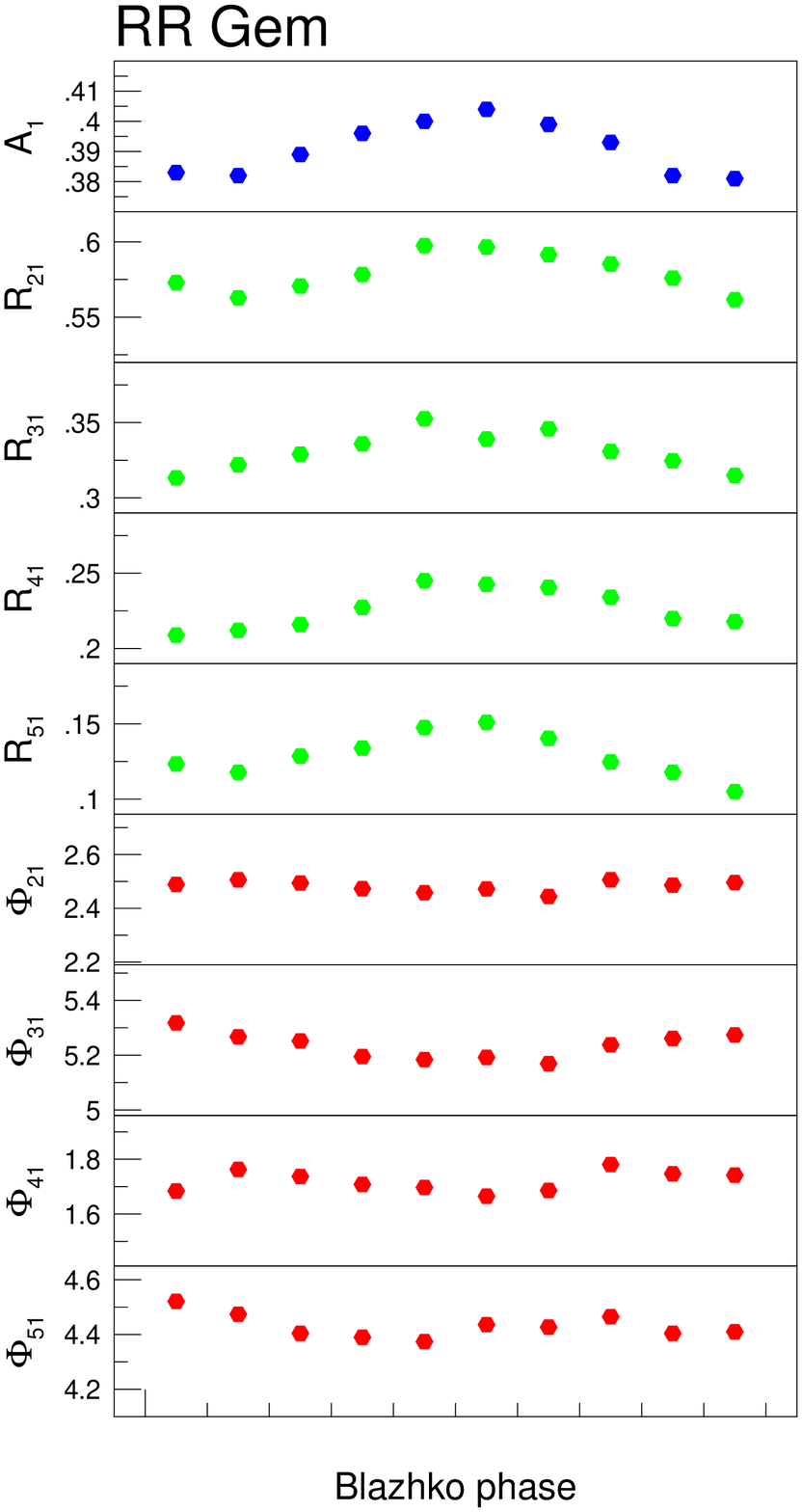}
\caption{Fourier parameters of the $V$ light curves in the different phases of
the Blazhko cycle of SS Cnc, and for comparison, of the other small amplitude,
short modulation period Blazhko star, RR Gem are plotted. $A1$ varies 0.02-0.03 mag 
during the Blazhko cycle in both of the plots (top panels), while the amplitude 
ratios ($R_{k1}$, 2.-5. panels) and phase differences ($\Phi_{k1}$, 6.-9. panels) 
show different patterns in the two stars. In SS Cnc the $R_{31}$ and $R_{41}$ 
amplitude ratios indicate variations which is not in phase with the variation of $A1$.
Parallel changes of the phase differences can be seen with increasing
amplitudes towards the higher orders: it is less than 0.1 rad in $\Phi_{21}$ but
reaches 0.4 rad in $\Phi_{51}$. This complex behavior of the changes in the
Fourier parameters reflects the compound, phase and amplitude modulation character 
of the Blazhko variation of SS Cnc. On the contrary, RR Gem exhibits much more
simple variations of its Fourier parameters.  Only the amplitudes show significant
changes, the amplitude ratios show similar amplitude, parallel variations with
$A1$. This may be the consequence of the dominance of its amplitude modulation and 
the symmetric behaviour of its amplitude and phase modulations as shown in 
Fig.~\ref{amp-phase}. 
\label{figfour}}
\end{figure}

The Fourier parameters of the pulsation light curves of the 10 data subsets 
for SS Cnc and RR Gem are plotted in Fig.~\ref{figfour}.  RR Gem shows the 
largest variation in the Fourier amplitudes and amplitude ratios ($R_{k1}=A_k/A_1$)
during the Blazhko cycle, while most of the Fourier parameters of SS Cnc vary
significantly. These properties reflect also that amplitude modulation is dominant 
in RR Gem, while in SS Cnc both amplitude and phase modulation have significant
amplitudes. An interesting feature of SS Cnc is that the variation in the amplitude 
ratios and the phase differences ($\phi_{k1}=\phi_{k}-k\phi_{1}$) are not harmonized 
in phase. It is worth noting that the amplitude of the variation increases towards 
the higher order phase differences.

As both SS Cnc and RR Gem were used as calibrating objects of the light 
curve--physical parameter relations of fundamental mode RR  Lyrae stars, we have
also checked how the detected changes in the Fourier parameters influence the
value of the derived physical parameters. The calculated [Fe/H] values in the 
different Blazhko phases vary by about 0.2  according to the formula which involves 
the period and $\phi_{31}$ \citep{jk96}. In the calculated absolute brightness values
\citep[using the formula derived for $Set A$ in][]{kw} there are less than 0.02 mag
differences for both of the stars. These values have the same  order of magnitudes  
if compared to the estimated accuracies of the formulae. We can thus conclude 
that small amplitude modulation of RRab stars probably does not influence severely 
the applicability of the light curve--physical parameter formulae.

It was already shown in \citet{rrg} that in contrast with theoretical expectations 
the amplitude ratios and the phase differences of the $kf_0+f_m$ and $kf_0-f_m$ 
modulation side lobe component pairs in the different colors and orders do not have 
the same value. The largest deviations of the four colors and the first five harmonic
orders are 0.25 in the amplitude ratios and $30\deg$ in the phase differences in RR
Gem. These differences are substantially larger, 3.1 and $76\deg$ in SS Cnc, 
respectively. There are slight, but significant differences between the results in 
the four colors of the same order, and unexpectedly large differences between the 
different order's data.

\subsection{Color changes}

\begin{figure}
\epsscale{1.1}
\plotone{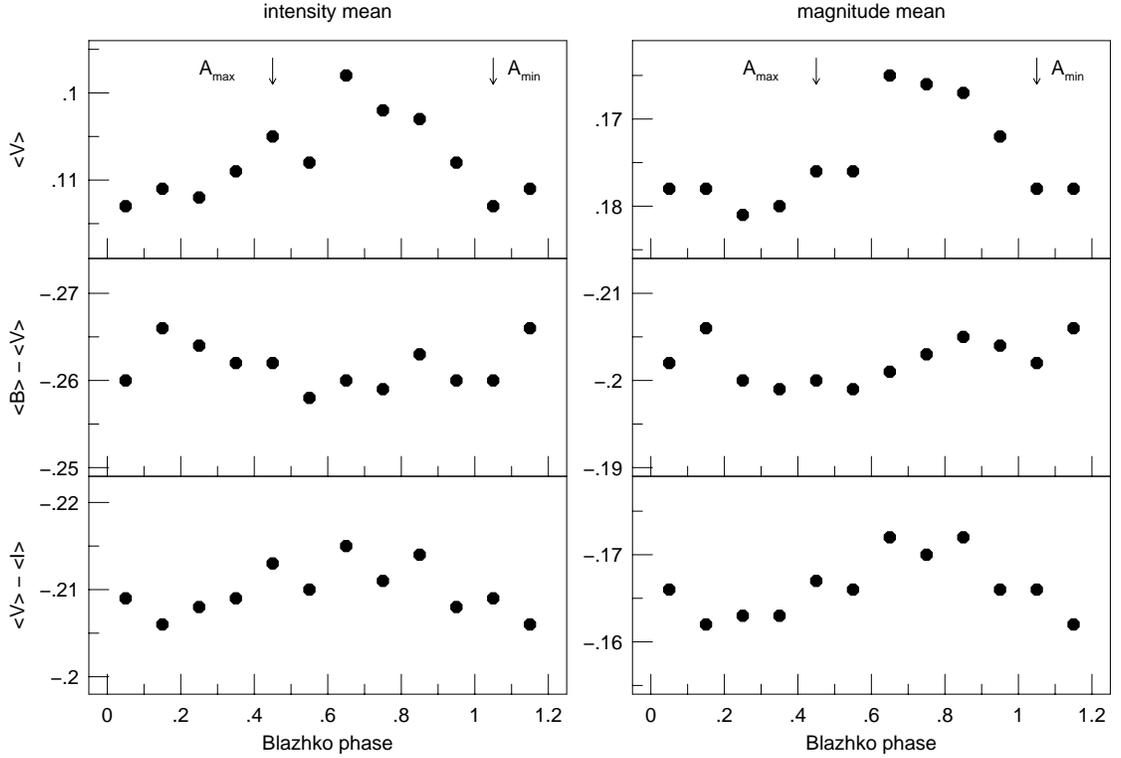}
\caption{Intensity and magnitude mean $\langle V \rangle$ brightness and $\langle B
\rangle - \langle V \rangle$ and $ \langle V \rangle - \langle I \rangle$ color 
changes during the Blazhko cycle of SS Cnc. Both intensity and magnitude mean 
$\langle V \rangle$ values indicate slight, $\sim0.01$ mag changes with minimum and 
maximum brightnesses preceding by $0.1$-$0.3$ in phase the minimum and maximum 
amplitude Blazhko phases. The color changes are even smaller than the detected 
changes in the mean brightnesses, they are of the order of millimagnitudes. 
The $\langle V \rangle - \langle I \rangle $ mean colors show slight parallel 
changes with the mean brightness variations, while the $\langle B \rangle - 
\langle V \rangle$ values do not show any definite trend.
\label{figszin}} 
\end{figure}

The intensity mean color changes during the Blazhko cycle indicate some tens of 
degrees mean effective temperature changes of RR Gem \citep{rrg}. 
The star seems to be slightly cooler in its smaller amplitude state than when its
pulsation amplitude is larger. Fig.~\ref{figszin} shows the magnitude and intensity 
mean $V, B-V$ and $V-I_C$ magnitudes and colors in the different phases of the 
Blazhko cycle of SS Cnc. The means are defined as the zero point term of the 
Fourier fits of the full pulsation light curves in each Blazhko phase interval. 
These values have been determined for the $B,V,R_C,I_C$ light curves in the 
different Blazhko phases, and the mean colors correspond to the differences of 
the mean magnitudes in the different wavelength bands. The accuracy of the mean 
magnitudes is estimated to be around 0.001-0.002 mag.

Both the intensity and magnitude mean $V$ brightnesses of SS Cnc vary during the
Blazhko cycle. Their amplitudes are about 0.012-0.015 mag. The brightest and
faintest states occur about 0.1-0.3 Blazhko phase after the phases of maximum and 
minimum amplitudes, respectively. The $\langle B \rangle-\langle V \rangle$ 
color index does not show systematic variation but its 0.005-0.007 mag scatter 
is larger than expected from the estimated accuracy of the data. The 
$\langle V \rangle-\langle I \rangle$ shows less than 0.01 mag amplitude variation
in line with the changes of the mean $V$ brightness.

\section{The discrepant properties of $f_m$}

It was a matter of debate for long whether the modulation frequency has 
observable amplitude in Blazhko stars at all. 
\citet{rvuma} and \citet{rsboo} detected 0.011 and 0.005 mag amplitudes 
of the modulation frequency in the $B$ data of the two large modulation amplitude 
Blazhko stars, RV UMa and RS Boo. During the modulation period the average 
brightness varies with an overall amplitude of 0.006 mag in 731 Blazhko
variables in the LMC \citep{machoab}. We have detected 0.006 and 0.008 mag 
$B$ amplitudes  of the modulation frequency of RR Gem and SS Cnc, respectively.
These results indicate that it is a general property of Blazhko stars that 
$f_m$ have similar amplitudes of the order of or less than 0.01 mag,
independently of the amplitude and other characteristics of the modulation.
This is in contrast with the large variety of the amplitudes of the $kf_0 \pm f_m$ 
modulation components, which exceed even 0.1 mag in some cases 
(see e.g. \citet{ahcam,arher}).

Another peculiarity of the $f_m$ frequency component is summarized in 
Table~\ref{tablerat}. This table lists the amplitude ratios and phase 
differences of the different color curve solutions for the modulation ($f_m$), 
and for the mean values of the pulsation ($kf_0$), and modulation 'side lobe' 
frequencies ($kf_0 \pm f_m$) for SS Cnc and RR Gem. There are no differences 
in the mean values of the amplitude ratios and phase differences of the pulsation 
and modulation side lobe frequency values of the two stars, they have the same 
values within the range of the scatter of the values for the different harmonic 
orders of the same star. The amplitude ratios detected for $f_m$ are, however, 
significantly different from these values. Towards redder colors the amplitudes 
of $f_m$ are much smaller for SS Cnc and much larger for RR Gem than detected 
for all the other frequency components. The $A_B/A_V$ amplitude ratios of $f_m$ 
are the less deviant, but for SS Cnc it is definitely smaller than the $A_B/A_V$ 
amplitude ratios of the other frequency components.

It is a natural expectation that the physics of the light variation 
determine the amplitude ratios in different colors. Therefore, the different 
color dependence of the amplitudes of $f_m$ and the similarity of the amplitude 
ratios of the pulsation and modulation side lobe frequencies indicate that  
$f_m$ has basically different physical origin than the radial mode pulsation has, 
while the side lobe frequencies share somehow the properties of the pulsation.
If nonlinear interaction between a millimagnitude amplitude signal ($f_m$) and the 
two orders of magnitude larger amplitude pulsation calls forth to the appearance 
of the side lobe frequencies, then the properties of the combination frequencies 
would more probably inherit the properties of the dominant pulsation frequencies 
than those of the small amplitude signal.  

This result if strengthened by detailed  observations of other Blazhko stars 
may lead to the conclusion that the real physically reasonable frequency of the
modulation is  in fact $f_m$. Any interpretation of the modulation which is 
connected to nonradial mode frequencies excited in the vicinity of the radial
mode  pulsation frequency may have difficulties to explain the discrepant behavior
of $f_m$, which might question the validity of the identification of any of the 
side lobe frequencies with a nonradial mode frequency.

\section{Conclusions}

Based on our photometric investigation of SS Cnc, and the previous 
study of RR Gem we add the following empirical facts to our knowledge 
of the modulation properties of RRab stars.

\begin{itemize}

\item{Small amplitude modulations with Fourier amplitudes of the modulation frequency
components of the order of millimagnitudes can be detected in RRab stars.}

\item{The amplitude decrease of the modulation components towards higher harmonic 
orders is significantly less steep than the amplitude decrease of the harmonic
components of the pulsation.}

\item{The changes of the Fourier parameters of the pulsation light curve 
in the different Blazhko phases are star by star different.}

\item{Based on the variations of the mean colors during the Blazhko cycle
the mean effective temperatures of the stars vary by less than $\sim 50$ K.}

\item{The color dependences of the amplitudes and phases of the pulsation and
modulation side lobe frequencies are the same in the different harmonic orders 
for the two small modulation amplitude stars, but the modulation frequency itself 
($f_m$) shows significantly different color behavior.}

\end{itemize}

These items, though still do not solve the mystery of the Blazhko phenomenon
add important contributions to our knowledge of the modulation properties of 
RRab stars, which have to be explained by any valid physical description.
As a concluding remark we also emphasize that the discovery of small amplitude 
modulations questions the validity of any previous statistics on the occurrence 
of the modulation. Based on accurate, detailed photometric data we also think 
that the current classification of the modulation into $\nu_1$ and Blazhko
types according to the amplitudes  of the modulation side lobe frequencies is 
just an artifact of noisy data and bad data sampling.

Up to now 4 galactic field Blazhko stars are known with modulation periods shorter
than 20 days: SS Cnc, RR Gem, AH Cam \citep{ahcam}, and CZ Lac \citep{inprep}.
It was shown in \citet{acta} that the modulation period only of the short
pulsation period RRab stars can be anomalously short. A common feature of these 
stars is that they are all metal rich, with metallicity close to the solar value. 
However, metallicity does not have a definite effect on the period of the modulation.
Metal rich, short pulsation period Blazhko stars with long modulation periods
do also exist, e.g., the modulation period of RS Boo is 530 days
($P_{puls}=0.38$ d, [Fe/H]$ = -0.2$).

An important result of our investigations is the demonstration of the discrepant
behavior of the modulation frequency, $f_m$ itself. The invariance of its amplitude 
in Blazhko stars and the discrepant color dependence of the amplitudes of this 
component indicate that the triggering mechanism of the light variations with the 
modulation period is basically different than that of the radial mode pulsation. 
The amplitudes of the side lobe modulation frequencies have the same color 
dependence as the pulsation frequencies have suggesting that these components are 
physically better connected to the radial mode.
 
The explanations of the Blazhko phenomenon involving nonradial modes do not give 
a clear picture yet. \citet{vh} have found many unstable low degree ($l=1,2$) 
nonradial $g$-modes of high radial order ($n>80$) in RR Lyrae models. These modes 
have frequencies in the vicinity of the frequencies of the observed radial modes. 
However, there is no indication in the literature, that the low frequency domain 
($f_m < 0.15$ c/d) would have been searched for unstable nonradial modes as well. 
The only model which explains the existence of the side lobe frequencies and is more 
or less in accordance with observations identifies these frequencies with rotation 
induced $m=\pm 1$ pairs \citep{nd}. This model cannot explain, however, the observed 
high amplitudes and the strong asymmetry of the triplets \citep{dm}.

The discrepant behaviour of the light variations with the modulation frequencies
urges theoretical investigations focusing on the possibility that short frequency 
nonradial modes in RR Lyraes stars similarly to $\gamma$ Dor variables, may also
exist.

\acknowledgments
{We would like to thank the referee, Horace Smith his useful comments which helped to
improve the lucidity of the paper. The financial support of OTKA grants T-043504,
T-046207 and T-048961 is acknowledged.}

\clearpage

\clearpage

\clearpage

\clearpage

\clearpage

\clearpage

\clearpage

\clearpage

\clearpage

\clearpage

\begin{deluxetable}{lcrrrrrrrr}
\tabletypesize{\scriptsize}
\tablecaption{Fourier amplitudes (amplitude ratios) and phases (phase
differences) of the detected frequencies in the $BVR_CI_C$ colors of SS Cnc
\label{tablefour}}
\tablewidth{0pt}
\tablehead{
\colhead{} & \colhead{frequency c/d} & \colhead{$A_V$ mag} &
\colhead{$\phi_V$ deg.} &
\colhead{$A_B/A_V$} &
\colhead{$\phi_B-\phi_V$} & \colhead{$A_V/A_R$} & \colhead{$\phi_V-\phi_R$} &
\colhead{$A_V/A_I$} & \colhead{$\phi_V-\phi_I$} }
\startdata
$f_m$       & 0.188350  & 0.00736 &   3.5 & 1.147 & $-$11.0 & 1.543 &     2.9 & 2.462 &     6.3 \\
$f_0-f_m$   & 2.533946  & 0.01491 & 332.3 & 1.298 &  $-$1.6 & 1.289 &     2.1 & 1.734 & $-$10.4 \\
$f_0$       & 2.722296  & 0.41668 & 169.3 & 1.370 &     2.5 & 1.273 &     3.3 & 1.678 &     8.9 \\
$f_0+f_m$   & 2.910646  & 0.00422 & 205.6 & 1.322 &  $-$5.5 & 1.311 &     2.2 & 1.737 & $-$20.1 \\
$2f_0-f_m$  & 5.256242  & 0.00796 & 257.0 & 1.180 &  $-$6.8 & 1.259 &     4.2 & 1.502 &  $-$1.0 \\
$2f_0$      & 5.444592  & 0.23932 & 115.5 & 1.342 &     0.4 & 1.250 &     0.6 & 1.627 &     2.1 \\
$2f_0+f_m$  & 5.632942  & 0.00546 & 172.9 & 1.158 &  $-$5.1 & 1.273 &     0.4 & 1.722 &  $-$2.9 \\
$3f_0-f_m$  & 7.978538  & 0.01290 & 191.9 & 1.384 &     1.7 & 1.287 &  $-$4.7 & 1.746 &     2.7 \\
$3f_0$      & 8.166888  & 0.14311 &  82.3 & 1.306 &  $-$0.0 & 1.235 &     0.0 & 1.589 &     0.4 \\
$3f_0+f_m$  & 8.355238  & 0.00336 & 105.0 & 1.381 & $-$10.7 & 1.091 &     2.7 & 1.965 &    19.5 \\
$4f_0-f_m$  & 10.700834 & 0.00723 & 150.7 & 1.313 &  $-$3.7 & 1.300 &  $-$4.4 & 1.449 &  $-$4.0 \\
$4f_0$      & 10.889184 & 0.09797 &  42.6 & 1.300 &     0.1 & 1.226 &  $-$0.6 & 1.577 &  $-$1.0 \\
$4f_0+f_m$  & 11.077534 & 0.00382 &  81.3 & 1.264 &     8.1 & 1.492 & $-$19.9 & 2.403 &     7.3 \\
$5f_0-f_m$  & 13.423130 & 0.00590 & 105.4 & 1.164 &  $-$2.2 & 1.143 &  $-$0.4 & 1.827 &     2.4 \\
$5f_0$      & 13.611480 & 0.05957 &   7.7 & 1.278 &     0.6 & 1.251 &     0.2 & 1.584 &     0.8 \\
$5f_0+f_m$  & 13.799830 & 0.00249 &  41.6 & 1.100 &  $-$0.9 & 0.984 &     1.5 & 1.705 &  $-$4.9 \\
$6f_0-f_m$  & 16.145426 & 0.00559 &  64.5 & 1.170 &     1.9 & 1.337 &  $-$3.9 & 1.775 &  $-$2.5 \\
$6f_0$      & 16.333776 & 0.04308 & 324.0 & 1.294 &  $-$0.9 & 1.240 &  $-$0.4 & 1.594 &  $-$0.5 \\
$6f_0+f_m$  & 16.522126 & 0.00256 & 348.2 & 1.371 &    15.2 & 1.882 &  $-$3.3 & 2.048 &    12.3 \\
$7f_0-f_m$  & 18.867722 & 0.00462 &  36.4 & 1.409 &  $-$0.1 & 1.467 &     5.6 & 1.668 &    10.7 \\
$7f_0$      & 19.056072 & 0.03283 & 287.5 & 1.263 &  $-$1.3 & 1.243 &     2.1 & 1.636 &     0.3 \\
$7f_0+f_m$  & 19.244422 & 0.00172 & 336.9 & 1.320 & $-$26.5 & 1.117 &  $-$1.0 & 1.564 &    11.9 \\
$8f_0-f_m$  & 21.590018 & 0.00364 & 355.7 & 1.275 &  $-$5.0 & 1.071 &     8.2 & 2.011 &    13.9 \\
$8f_0$      & 21.778368 & 0.02139 & 248.1 & 1.262 &     0.2 & 1.280 &     2.4 & 1.695 &     1.4 \\
$8f_0+f_m$  & 21.966718 & 0.00102 & 261.8 & 1.539 &     4.9 & 1.457 &    21.5 & 1.172 &     2.7 \\
$9f_0-f_m$  & 24.312314 & 0.00278 & 308.1 & 1.068 &  $-$4.2 & 1.187 &     6.3 & 2.122 &  $-$1.0 \\
$9f_0$      & 24.500664 & 0.01513 & 198.3 & 1.294 &     0.7 & 1.192 &     0.2 & 1.589 &  $-$1.8 \\
$9f_0+f_m$  & 24.689014 & 0.00162 & 231.2 & 1.562 &    20.7 & 1.514 & $-$11.1 & 1.421 & $-$11.7 \\
$10f_0-f_m$ & 27.034610 & 0.00200 & 266.9 & 1.305 & $-$14.7 & 1.250 &     2.6 & 1.563 &    21.1 \\
$10f_0$     & 27.222960 & 0.01126 & 162.4 & 1.283 &  $-$4.7 & 1.169 &     1.9 & 1.532 &     2.8 \\
$11f_0-f_m$ & 29.756906 & 0.00135 & 206.1 & 1.689 &    13.6 & 0.849 & $-$14.5 & 1.262 &  $-$8.4 \\
$11f_0$     & 29.945256 & 0.00842 & 114.7 & 1.290 &  $-$7.5 & 1.168 &  $-$0.1 & 1.815 &     9.8 \\
$12f_0-f_m$ & 32.479202 & 0.00167 & 185.4 & 1.275 &  $-$1.5 & 0.898 &     5.9 & 1.653 &    23.0 \\
$12f_0$     & 32.667552 & 0.00621 &  64.6 & 1.417 &     1.3 & 1.250 &     2.7 & 1.451 &     1.3 \\
$13f_0-f_m$ & 35.201498 & 0.00105 & 127.3 & 1.486 &     0.2 & 0.938 &     1.1 & 1.522 &    33.6 \\
$13f_0$     & 35.389848 & 0.00534 &  19.8 & 1.356 &  $-$3.1 & 1.236 &     0.6 & 1.745 &     1.2 \\
$14f_0-f_m$ & 37.923794 & 0.00121 &  87.1 & 1.438 &    19.9 & 1.322 & $-$13.1 & 1.080 &  $-$2.4 \\
$14f_0$     & 38.112144 & 0.00402 & 332.4 & 1.313 &  $-$4.0 & 1.371 &     6.6 & 1.558 &     6.8 \\
$15f_0-f_m$ & 40.646090 & 0.00162 &  56.1 & 1.117 & $-$10.2 & 1.443 &    10.7 & 2.492 &    19.4 \\
$15f_0$     & 40.834440 & 0.00351 & 281.7 & 1.291 &     0.7 & 1.167 &    10.2 & 1.469 &     3.6 \\
$16f_0$     & 43.556736 & 0.00333 & 243.1 & 1.300 & $-$10.7 & 1.140 &     6.3 & 1.617 &    13.8 \\
$17f_0$     & 46.279032 & 0.00281 & 181.8 & 1.580 &     6.5 & 1.373 &     3.5 & 1.813 &    21.6 \\
$18f_0$     & 49.001328 & 0.00250 & 146.9 & 1.488 &  $-$2.1 & 1.471 &  $-$3.0 & 1.736 &    11.6 \\
$19f_0$     & 51.723624 & 0.00192 &  96.7 & 1.651 &    12.3 & 0.905 &  $-$1.3 & 1.049 &    15.0 \\
$20f_0$     & 54.445920 & 0.00280 &  55.8 & 1.168 &     7.0 & 1.544 &     3.5 & 1.818 &    20.8 \\
$21f_0$     & 57.168216 & 0.00221 &  19.3 & 0.986 &     3.3 & 1.435 &    14.7 & 2.105 &    20.3 \\
$22f_0$     & 59.890512 & 0.00158 & 324.1 & 1.994 &     5.6 & 1.047 &     8.3 & 1.975 &    26.9 \\
$23f_0$     & 62.612808 & 0.00181 & 288.1 & 1.039 &     1.1 & 1.275 & $-$10.4 & 1.926 &    35.4 \\
$24f_0$     & 65.335104 & 0.00206 & 263.4 & 1.087 &     0.1 & 1.537 &    10.5 & 2.512 &     8.9 \\
$25f_0$     & 68.057400 & 0.00134 & 213.7 & 1.224 &     7.8 & 1.164 &    24.3 & 1.207 &    19.1 \\
\enddata                                                    
\end{deluxetable}
\clearpage

\begin{deluxetable}{lrcrcrcrcrcrc}
\tabletypesize{\small}
\rotate
\tablecaption{Amplitude ratios and phase differences of the
 modulation ($f_m$), pulsation
($kf_0$),
and modulation 'side lobe' frequencies ($kf_0 \pm f_m$) 
of the different color light curve solutions for 
SS Cnc and RR Gem.
\label{tablerat}}
\tablewidth{0pt}
\tablehead{
\colhead{Star} & \colhead{$A_B/A_V$}&\colhead{$s.dev.$} & 
\colhead{$\phi_B-\phi_V$} &\colhead{$s.dev.$}& \colhead{$A_V/A_R$}&\colhead{$s.dev.$} 
& \colhead{$\phi_V-\phi_R$}&\colhead{$s.dev.$} &
\colhead{$A_V/A_I$}&\colhead{$s.dev.$} & \colhead{$\phi_V-\phi_I$}&\colhead{$s.dev.$}
}
\startdata
\sidehead{$f_m$}
SS Cnc& 1.15& &$-11.0$ && 1.54 && 2.9 && 2.46& & 6.3&\\
RR Gem& 1.29& &$-9.8$  && 1.13 && 9.8 && 1.07& & 1.6&\\
\sidehead{$kf_0$\tablenotemark{a}}
SS Cnc& 1.30 &0.03 &$-0.9$ &2.8 & 1.23 &0.04& 0.9 &1.3 & 1.63& 0.08 & 2.1& 3.8\\
RR Gem& 1.32 &0.07 &$-0.7$ &1.8 & 1.27 &0.07& 0.5 &1.9 & 1.67& 0.17 & 1.9& 3.6\\
\sidehead{$kf_0 \pm f_m$\tablenotemark{a}}
SS Cnc& 1.29 &0.08 &$-2.6$ &3.6 & 1.28 &0.02& $-0.7$& 4.5 & 1.61& 0.15 & $-3.2$& 5.5\\
RR Gem& 1.36 &0.07 &$-1.1$ &4.1 & 1.22 &0.13& $-2.5$& 7.1 & 1.55& 0.11 & $-7.7$& 6.7\\
\enddata
\tablenotetext{a}{Mean values and standard deviations of the amplitude ratios and
phase differences of those pulsation and modulation side lobe frequency components 
that have $V$ amplitudes larger than or nearly equal to the amplitude of $f_m$ 
are given. This means 11 and 12 pulsation and 4 and 6 modulation frequency
components for SS Cnc and RR Gem, respectively.} 
\end{deluxetable}

\end{document}